\documentstyle[12pt]{article}
\begin{document}

\setcounter{page}{0}
\thispagestyle{empty}

{}~\vspace{2cm}

\begin{center} {\Large {\bf THE PHENOMENOLOGY OF SCALAR COLOUR
      OCTETS}} \\
 \vspace{1.5cm}
 {\large
  N.V.Krasnikov\footnote{
On leave of absence from Institute for Nuclear Research, Moscow
117312, Russia.\\
e-mail: KRASNIKO@MS2.INR.AC.RU} \\
 \vspace {0.5cm}
 {\em Laboratoire Physique Theorique ENSLAPP\\
LAPP, B.P. 110, F-74941, Annecy-le-Vieux Cedex, France}}
\end{center}

\vspace{1.5cm}\noindent
 \begin{center} {\bf Abstract} \end{center}

\date{June 1995}
We discuss the phenomenology of colour scalar octet particles. Namely,
we consider the discovery potential of scalar octets at LEP, FNAL
and LHC. Octet scalars decay mainly into two gluino and new hadrons
composed from scalar coloured octets are rather longlived
($\Gamma \leq  O(10)Kev$). For the scalar octet masses($M \geq O(50)Gev$)
the scalar octets producing at FNAL or LHC will decay into two gluons
that
leads to additional four-jet events. We propose to look for the scalar
octets by the measurement of the distributions of the four-jet
differential
cross section on the invariant two-jet masses. Scalar octets naturally
arise in models with compactification of additional dimensions. In such
models the branching ratio for the scalar octet bound state into 2
photons is $O(10^{-2})$ that leads to the events with two photons and
two jets. We also point out that the
current experimental data don't contradict to the existence of light
($M \sim O(1)Gev$) scalar octets. Light scalar colour octets give
additional
contribution to the QCD $\beta$ -function and allow to improve agreement
between deep inelastic and LEP data.

\newpage

The aim of this note is the discussion of the phenomenology of scalar
colour octets. The relatively light ($M \leq O(1)Tev$) scalar colour
octets are predicted in some nonsupersymmetric and supersymmetic GUTs
\cite{1,2}. Here we consider the discovery potential of scalar octets at
LEP, FNAL and LHC. Colour octet scalars decay mainly into two gluons
with rather small decay width ($\Gamma \leq O(10)Kev$), so new hadrons
composed from colour scalar octets are longlived. The scalar octets producing
at FNAL or LHC (for scalar octet masses $M \geq O(50)Gev $)  decay
into two gluino that leads to additional four-jet events. We propose
to look for scalar octets by the measurement of the distributions of the
four-jet differential cross section  on the invariant two-jet masses. At
least, the accurate measurement of the four-jet differential cross section
allows to extract the lower bound for the octet scalar cross section.
Scalar octets naturally arise in models with compactification of
additional dimensions. In such models the branching ratio for the scalar
octet bound state into two photons is
$Br(\Phi g \rightarrow \gamma \gamma) \sim O(10^{-2})$. For such
models we shall have the events with two photons and two jets that
allows to detect scalar octets using this mode.
We also point out that the current experimental data don't contradict to the
existence of light ($M \sim O(1)Gev$) scalar octets. Light scalar octets give
additional contribution to the QCD $\beta$-function and allow to improve
agreement between deep inelastic and LEP data.

To be precise in this paper we consider colour light scalar octets neutral
under $SU(2) \otimes U(1)$ electroweak gauge group. Such particles are
described by the selfconjugate scalar field $\Phi^{\alpha}_{\beta}(x)$
 ($(\Phi^{\alpha}_{\beta}(x))^{+} = \Phi^{\beta}_{\alpha}(x),  \,
\Phi^{\alpha}_{\alpha}(x) = 0 $)  interacting only with gluons. Here
$\alpha = 1,2,3  ; \,\beta = 1,2,3$ are $SU(3)$ indices.  The scalar potential
for the scalar octet field $\Phi^{\alpha}_{\beta}(x)$ has the form

\begin{equation}
V(\Phi) = \frac{M^2}{2}Tr(\Phi^{2}) + \frac{\lambda_{1}M}{6}Tr(\Phi^{3}) +
\frac{\lambda_{2}}{12}Tr(\Phi^4)+ \frac{\lambda_{3}}{12}(Tr\Phi^{2})^2
\end{equation}

The term $\frac{\lambda_{1} M}{6}Tr(\Phi^3)$ in the scalar potential (1)
breaks the discrete symmetry $\Phi \rightarrow -\Phi$. The existence of such
term in the lagrangian leads to the decay of the scalar octet mainly into
two gluons through one-loop diagrams similar to the corresponding one-loop
diagrams describing the Higgs boson decay into two photons. One can find that
the decay width of the scalar octet is determined by the formula
\begin{equation}
\Gamma(\Phi \rightarrow g\,g) = \frac{15}{4096\pi^{3}}\alpha_{s}^{2}c^2
\lambda_{1}^{2}M \,,
\end{equation}
where

\begin{equation}
c = \int_{0}^{1} \int_{0}^{1-w} \frac{wu}{1-u-w}\,du\,dw = 0.048
\end{equation}
and $\alpha_{s}$ is the effective strong coupling constant at some
normalization point $\mu \sim M_{z}$.
Numerically for $\alpha_{s} = 0.12$ we find that
\begin{equation}
\Gamma(\Phi \rightarrow gg) = 0.39\cdot 10^{-8}\lambda_{1}^{2}M
\end{equation}

 From the requirements that colour $SU(3)$ symmetry is unbroken (the minimum
$\Phi^{\alpha}_{\beta}(x) = 0 $ is the deepest one) and the effective
coupling constants $\overline{\lambda}_{2}$ , $\overline{\lambda}_{3}$
don't have Landau pole singularities up to the energy $M_{0} = 100\cdot M$
we find that $\lambda_{1} \leq O(1)$ . Therefore, the decay width of
the scalar colour octet is less than $O(10)ev, O(100)ev, O(1)Kev,
O(10)Kev$ for the octet masses $M = 1,\, 10,\, 100,\, 1000$ Gev
correspondingly. It means that new hadrons composed from scalar octet
$\Phi$, quarks and gluons ($\overline{q} \Phi q, \,\Phi g, \,qqq\Phi $ are
longlived even for very high scalar octet mass.

Light scalar octets with the mass $M = O(several) \,Gev$ give additional
contribution to the QCD $\beta$-function that changes evolution
of the QCD effective strong coupling constant. For instance, in one-loop
approximation we have the following formula \cite{3,4} for the effective
strong coupling constant:
\begin{equation}
\alpha_{s}(Q) = \frac{4\pi}{b_{0}ln(Q^{2}/\Lambda^2)} \,,
\end{equation}

where $b_{0}=11-\frac{2}{3}N_{f}$ for the standard case and
$b_{0}= 11-\frac{2}{3}N_{f} +\frac{1}{2}$ for the case when we take
into account light scalar octet in the loop. In complete analogy with the
case of light gluino \cite{5,6} an account of light scalar octets leads
to the modification of the strong coupling constant at the
$M_z$ scale extracted from low energy deep inelastic data and data on the
$\tau$-lepton decay width. Namely, one can find that the "modified"
strong coupling constant taking into account the QCD evolution due to
light scalar octet is
\begin{equation}
\alpha^{mod}_{s}(M_z)= \frac{1}{\frac{1}{\alpha_{s}(M_{z})}-
0.08ln(\frac{M_{z}}{M})}
\end{equation}

For instance, for $\alpha_{s}(M_z) = 0.113$ extracted from deep inelastic
data for $Q_{0} = 5 Gev$ we find that
$\alpha^{mod}_{s}(M_z) = 0.1161;\, 0.1153;\, 0.1146;\, 0.1136$ for the scalar
octet masses $M = 5; \,10;\, 20;\, 30 \, Gev$ correspondingly. So we find that
the maximal effect is the increase of the effective strong coupling constant
by 0.003 that is welcomed effect since it improves the agreement
between strong coupling constant extracted from Z-boson total hadronic
decay width and deep inelastic data. For the current situation with
experimental determination of $\alpha_{s} (M_z)$ see ref.\cite{7}.

Consider now the possibility of the discovery of scalar octets at LEP1.
For the scalar octets lighter than $\frac{M_z}{2}$ the differential
decay width
\begin{equation}
Z(p) \rightarrow \overline{q}(p_1)q(p_2)\Phi (p_3) \Phi(p_4)
\end{equation}

in the leading approximation on strong coupling constant $\alpha_{s}(M_z)$
for massless quarks is determined by the formula
\begin{equation}
d\Gamma(Z \rightarrow \overline{q}{q} \Phi \Phi)\cdot (\Gamma (Z
\rightarrow hadrons))^{-1} = A\,dm_{12}^{2}dp^{2}\,,
\end{equation}

\begin{equation}
A = \frac{4\alpha_{s}^{2}}{3\pi^{2}p^2}(1-\frac{4M^2}{M^2_z})^{\frac{3}{2}}
\cdot [BC^{-1}ln(\frac{C+D}{C-D}) - 2D]
\end{equation}
Here $\alpha_{s}$ is the effective strong coupling constant at some
normalization point $\mu \sim M_z$  , $ m_{12}^{2} = (p_{1} + p_{2})^2, \,
\, p^2 = (p_{3} + p_{4})^{2} \,\,,C =\frac{1}{2}(M_z^2 + p^2 - m_{12}^{2})\,\,,
B = C^2 + \frac{1}{2} m_{12}^{2}(M_z^2 + p^2)\,\,,
D^2 = C^2 - M_{z}^{2}p^{2}\,  $ .
In our numerical estimates we shall take $\alpha_{s} = 0.12$.
For the branching ratio $B = 10^{3} \Gamma (Z \rightarrow \overline{q} q \Phi
\Phi)
\cdot(\Gamma(Z \rightarrow hadrons))^{-1}$ our numerical
results are presented in table 1.
For light scalar octets ($M \sim O(several)\, Gev$) the process
$Z \rightarrow \overline{q}q \Phi \Phi$ qives additional contribution to the
standard QCD four-jet production in Z-decay. We have found that the
four-jet cross section of the process $Z \rightarrow \overline{q} q \Phi \Phi$
is approximately 15\%  of standard QCD four-jet cross section
$Z \rightarrow \overline{q}q \overline{q}^{'}q{'}$
which in turn is around 5\% of the total four-jet cross section
\cite{8}. So the discovery at LEP of light scalar octets by the
measurement of the four-jet cross section is rather problematic.
For the scalar masses $M \geq O(10)Gev$ the scalar octet decaying
into two gluons produces two gluon jets, so we shal have 6 jet events
with 4 gluon jets. Unfortunately we don't know rather well 6 jet
cross sections. We can estimate that 6 jet cross section
is $\alpha^{2}_{s} \sim 0.01 $ smaller than
4-jet cross section, so the standard 6 jet QCD cross section
is of the same order of magnitude as (at least for $M \sim 10 Gev$) the
6 jet cross section due to the scalar octet decays. By the measurement of
the differential 6 jet cross section
\begin{equation}
\frac{d\sigma^{2}}{dm_{12}dm_{34}}
\end{equation}
(here $m_{12}^{2} = (p_{1,jet}+p_{2,jet})^2$ and
$m_{34}^{2} = (p_{3,jet}+p_{4,jet})^2$
are the invariant two-jet square masses) it is possible to discover
the scalar octets with masses $ 10 Gev \leq M \leq 20 Gev$
provided the accuracy in the determination of the two-jet invariant mass
is less than 10\% since in this case we earn additional factor $ O(100)$
for the suppression of the background.
 So, it would be very interesting to consider the distributions
of 6 jet events at LEP1 on the two-jet invariant masses. As I know
there was not such analysis of LEP1 data.

Consider now the production of scalar octets at FNAL and LHC. The
corresponding lowest order predictions for the parton cross sections
have the form
\begin {equation}
\frac{d\sigma}{dt}(\overline{q}q \rightarrow \Phi \, \Phi) =
\frac{4\pi\alpha^{2}_{s}}{s^4}(tu - M^4) \,,
\end{equation}
\begin{eqnarray}
\frac{d\sigma}{dt}(gg \rightarrow \Phi \Phi) &=& \frac{\pi\alpha_{s}^{2}}
{s^2}(\frac{7}{96} + \frac{3(u-t)^2}{32s^2})(1 + \frac{2M^2}{u-M^2} +
\frac{2M^2}{t-M^2} +  \\ \nonumber
&&\frac{2M^4}{(u-M^2)^2} + \frac{2M^4}{(t-M^2)^2} +
\frac{4M^4}{(t-M^2)(u-M^2)})\,,
\end{eqnarray}
\begin{equation}
\sigma(\overline{q}q \rightarrow \Phi \Phi) = \frac{2\pi\alpha^{2}_{s}}{9s}
k^{3}\,,
\end{equation}
\begin{equation}
\sigma(gg \rightarrow \Phi \Phi) = \frac{\pi\alpha^{2}_{s}}{s}(\frac{15k}{16}
+\frac{51kM^2}{8s} + \frac{9M^{2}}{2s^2}(s-M^{2})ln(\frac{1-k}{1+k})) \,,
\end{equation}
where $k = (1 - \frac{4M^{2}}{s})^{\frac{1}{2}}$. We have calculated the
cross sections for the production of scalar octets at FNAL and LHC using
the results for the parton distributions of ref.\cite{9}, namely,
in our calculations we have used set 1 of the parton distributions.
We have found that at LHC the main contribution
($\geq $ 95\%) comes from the gluon annihilation into two scalar octets
$gg \rightarrow \Phi \Phi $, whereas at FNAL gluon-gluon and
quark-antiquark annihilation cross sections are comparable.
The results of our calculations are presented in tables 2 and 3.
For light scalar octets two gluon and quark-antiquark annihilations
into two scalar octets give additional contribution to the two-jet
cross section. However, this additional contribution is rather small.
For instance, the cross section for gluon-gluon scattering is \cite{10}
\begin{equation}
\frac{d\sigma}{dt}(gg \rightarrow gg) = \frac{9\pi\alpha^{2}_{s}}{2s^2}
[3 - \frac{tu}{s^2} - \frac{su}{t^2} - \frac{st}{u^2}]
\end{equation}
Even for the most favorable case $t=u=-\frac{s}{2}$  the cross section
(12) is 20 times less than gluon-gluon cross section (15). So the perspective
to detect light scalar octets by the measurement of the two-jet cross sections
looks hopeless. For rather big values of the scalar octet mass ($M \geq
O(50)Gev$) the scalar octets decay into two gluons that leads to the
four-jet events. The cross section for the scalar octet production
for $M \sim 100 Gev$ is $O(10^{-3})$ of the standard QCD two-jet
cross section and it is (10 - 100) times smaller than the standard
4-jet QCD cross section. So by the measurement of the two-jet invariant
masses like as in the case of LEP1 it is possible (if we know two-jet
invariant masses with an accuracy better than 10\%) to earn additional
factor $\sim 100$ and to discover or at least to obtain lower bound on the
scalar octet cross section for scalar octet masses $ O(several)Gev \leq
M \leq O(200) Gev$ at LHC.

It should be noted that scalar octets  naturally arise in models with
compactification of additional space dimensions \cite{11}. As a toy
model consider 5-dimensional QCD with massless quarks. Let us
compactify the 4 th coordinate $x_{4}$ on the torus, i.e. impose the
boundary conditions
\begin{equation}
A_{M}(x_{\mu}, x_{4}) + R_{c}) = A_{M}(x_{\mu}, x_{4}) ,
\end{equation}

\begin{equation}
\Psi(x_{\mu}, x_{4} + R_{c}) = \Psi(x_{\mu}, x_{4})
\end{equation}
After compactification we obtain at tree level  massless gluon
field $A_{\mu}(x_{\mu})$ , massless scalar octet
$\Phi(x_{\mu}) =  A_{4}(x_{\mu})$, massless quarks plus the
infinite tower of massive excitations of gluons and quarks with
masses proportional to the inverse compactification radius
$R_{c}^{-1}$  . At quantum level massless octet $\Phi(x_{\mu})$
acquire a mass $ M \sim (O(\frac{\alpha_{s}}{2\pi}))^{0.5}R_{c}^{-1}$.
So, for the models with big compactification radius
$R_{c}^{-1} \leq 10 Tev$ we expect that the scalar octet mass is less
than $O(1) Tev$ . Due to the interaction of scalar octet with
the excitations of quarks scalar octet will decay at one loop level
mainly into two gluons, however the branching ratio of the colourless
bound state of scalar octet and gluons into two photons is not small
and one can estimate that it is
$Br(\Phi g \rightarrow \gamma \gamma ) \sim O((\frac{\alpha_{em}}
{\alpha_{s}})^{2}) \sim O(10^{-2})$. So we shall have events with
two jets and two photons (one of the producing octet bound state
decays into two gluons and the other one decays into two photons).
Therefore the search for the Higgs boson at LHC using two photon
mode is simultaneously the search for the scalar octets, however for the
scalar octet case besides two photons we shall have in addition two
jets that makes the search for the scalar octets much more easier
than the search for the Higgs boson. For instance, for the integral
luminosity $10^{5} (pb)^{-1}$ we expect for the scalar octet mass
$M = 200 Gev$  and $M = 700 Gev $ $O(10^{6})$ and $O(10^{3})$ events
with two photons and two jets correspondingly , so it would be
possible to discover at LHC the scalar octets with masses lighter
than 1 Tev.

To conclude, in this note we have studied the perspectives of the discovery
of scalar octets at LEP1, FNAL and LHC. New hadrons composed from scalar
octets are rather longlived even for high scalar octet masses. We have
found that the existence of light scalar octets with the masses
$O(several)\, Gev$ don't contradict to the existing experiments.
Heavy scalar octets could be discovered at LHC, FNAL or LEP by the
measurement of the distributions of the differential cross sections
on the invariant two-jet masses. Scalar octets naturally arise in
models with compactification of additional space dimensions. In such
 models the branching ratio of scalar octet and gluon
colourless bound state
into two photons is $O(10^{-2})$  so scalar octets
 in such models could be detected by the search for the events with
two photons and two jets.

I am indebted to the collaborators of LAPP theoretical department for
the hospitality during my stay at Annecy where the final text of this
note has been written. I am pleased to  thank the collaborators of the
INR theoretical and experimental departments and especially to
S.Gninenko for discussions and critical comments. The research
described in this publication was made possible in part by
Grants N6G000, N6G300 from the International Science Foundation and
by Grant 94-02-04474-a of the Russian Scientific Foundation.

\newpage

Table 1. The branching $B = 10^{3}\cdot\Gamma(Z \rightarrow
\overline{q}q\Phi\Phi)\cdot ((\Gamma Z \rightarrow hadrons))^{-1})$
for different scalar octet masses

\begin{center}
\begin{tabular}{|l|l|l|l|l|l|l|l|l|}
\hline
M(Gev) & 2 & 5 & 10 & 15 & 20 & 25 & 30 & 35 \\
\hline
B & 7.25 & 1.63 & 0.22 & 0.037 & 0.0058 & 0.00078 &
 $7.0\cdot10^{-5}$ & $2.7\cdot10^{-6}$   \\
\hline
\end{tabular}
\end{center}

Table 2. The cross section $\sigma(p p \rightarrow \Phi \Phi
\, + \, ...)$ in pb for different values of octet masses and normalization
point$\mu$ at LHC

\begin{center}
\begin{tabular}{|l|l|l|l|l|l|l|l|l|}
\hline
M(Tev) & &  2 & 1.5 & 1 & 0.75 & 0.5 & 0.3 & 0.2 \\
\hline
$\sigma$ & $\mu = 3.75Tev$ & 0.00024 & 0.004 & 0.093 & 0.60 & 6.0 & 76.3 &
786.4 \\
\hline
$\sigma$ & $\mu = 2M$ & 0.00026 & 0.0042 & 0.10 & 0.69 & 7.40 & 84.3 & 701.4 \\
\hline
$\sigma$ & $\mu=4M$ & 0.00028 & 0.0043 & 0.092 & 0.619  & 6.6 & 91.0 & 832  \\
\hline
\end{tabular}
\end{center}

Table 3. The cross section $\sigma(\overline{p} p \rightarrow \Phi \Phi \, +
\, ...)$ in pb for different values of octet masses and normalization point
$\mu$  at FNAL.

\begin{center}
\begin{tabular}{|l|l|l|l|l|l|l|l|l|l|}
\hline
M(Gev) & & 300 & 250 & 200 & 150 & 125 & 100 & 75 & 50 \\
\hline
$\sigma$ & $\mu=450 Gev$ & 0.014 & 0.07 & 0.40 &3.0 & 9.9 & 38.8 & 198.4 & 1620
  \\
\hline
$\sigma$ & $\mu=2M$ & 0.013 & 0.074 & 0.42 & 3.56 & 10.8 & 52.4 & 267.4 & 2940
\\
\hline
$\sigma$ & $\mu = 4M$ & 0.014 & 0.058 & 0.33 & 2.8 & 9.6 & 40.3 & 208.3 & 1782
\\
\hline
\end{tabular}
\end{center}

\end{document}